# COLLABORATION FOR ENHANCING THE SYSTEM DEVELOPMENT PROCESS IN OPEN SOURCE DILIGENCE.

Murtaza Hussain Shaikh

Department of Computer and Information Science, Norwegian University of Science and Technology (NTNU), Trondheim – Norway.  Phone: +4796826676

Shaikh@stud.ntnu.no

## ABSTRACT

*According to different opponents and commercial giants in software industries, the open source style software development has enough capacity to complete successfully the large scale projects. But we have seen many flaws and loops in collaboration and handling of mega scale projects in open source environment. Perhaps the collaboration is a key of successful project development. In this article we have tries to identify different feasible and reliable solution to a better collaboration ways in the open source system development. Some of the issues also that are found in the development phase of the open source have been identified and a proposed solution by explaining Successful communities such as GNU, the Apache Software Foundation, and Eclipse Foundation is discusses in this research article.  It must be kept in mind that to improvement the collaboration in open source environment both the development community and the people should be more creative.*

## 1. INTRODUCTION

As the computer science field has grown to be ubiquitous part of human`s life the need for making s software and hardware support to fit the needs of users is constantly increasing. Open source development projects are generally internet base communities of system developers. The systems that are developed made freely available to all that adhere to the licensing terms specified by the open source project. The open source system projects and their development processes have spread rapidly and widely, and many thousands exist in today's era. The number of developers and communities participating in each project ranges from a few too many thousands and the numbers of users of the software produced by open source systems development projects range from few too  many millions [1].  Well known examples of open source systems having many users are LINUX / UNIX operating systems, Apache server and Perl Programming language.

The phenomenon of open source systems with better development process is interesting. Its functionalities in the market together with its novel modes of operation and security pose major and exciting questions regarding lot of uncertainties & risks today concerning our privacy and trust, principles by which productive work can best be organized in secure manner. There are different collaborative solutions that are proposed on these special issues as a form for some exciting new work and voices on open source systems [1].

## 2. THE OPEN SOURCE DILIGENCE

Today in software industry, firms, individuals appropriate and secure financial rewards from their packaged software products by two main principle means. The first is software industry uses licensing arrangements based on copyrights law. A software license provides individuals and group with the legal rights to use a piece of software often in return for a licensing fee [2]. In current commercial practice, most software is licensed rather than sold to a third party as intellectual property. A software license typically restricts the numbers of computers the software can run on, the number of software users, backups, and simultaneous use of backups. Second, many commercial software firms protect the software's *"source code"*, which is a sequence of instructions to be executed by a computer to accomplish a program's purpose. Programmers write software in the form of source code, and also *"Document"* that source code with brief written explanations of the purpose and design of each section of their program [1]. In the early days of computer programming such commercial, protected, and packaged software was a rarity; if you wanted a particular program for a particular purpose you typically wrote the code yourself or hired it done. Much of the software development in 1970`s was carried out in academic and corporate laboratories by scientists and engineers [14].

These individuals found it a normal part of their research culture to freely give and exchange software they had written, to modify and build upon each other's software both individually and collaboratively, and to freely give out their modifications in turn. The free software idea did not immediately become main stream, and industry was especially suspicious of it. It is agreed in [14] that a significant part of the problem resided in Stallman's term *"free"* software; which might understandably have an ominous ring to the ears of business people. Accordingly they, along with other prominent hackers, founded the *"open source"* software movement.

## 3. PROBLEM DESCRIPTION

OS (open source) software development projects are typically initiated by an individual or a small group with an idea for something interesting .They themselves want for an intellectual or business reason. In [4] the author suggests *"Every good work of software starts by scratching a developer's personal itch, software developers spend their days grinding away for pay at programs they neither need nor love"*. The source code for this first version is then made freely available to all via downloading from an Internet website established by the project [1]. In the case of projects that are successful in attracting interest, others do download and use the code and some of these do go on to create new and modified code based on their own interests. Most then post what they have done on the project website for use and critique by anyone who is interested.

Information systems are any combination of information technology and people's activates using that core technology to support management and decision-making [5]. The scope of information systems is pervasive; it extends to need people in different sectors apart from the commercial business settings. One of the fields which information systems can support is the open source systems in collaborative support environment. The main focus of this article is to understand *"how the open source systems can be useful to automate in computer support & collaboration and what is the basic terminology that plays a vital role in the Collaboration"?*

## 4. SYSTEM`S TOPOGRAPHY

To a large scale, there are many features of open source systems, but few are as described as below;

## 4.1. Free Redistribution

Licensing of the open source software system shall not restrict any party from selling it or giving any of its components [8].

## 4.2. Source Codes

The program must include source code, and must allow the distribution in source code as well as compiled form. Where some form of a product is not distributed with source code, there must be a well-publicized means of obtaining the source code for no more than a reasonable reproduction cost preferably, downloading via the Internet without charge [8].

## 4.3. Derived Works

The license must allow modifications and derived works, and must allow them to be distributed under the same terms as the license of the original software [8].

## 4.4. Integrity in Codes

The license may restrict source-code from being distributed in modified form only if the license allows the distribution of *"patch files"* with the source code for the purpose of modifying the program at build time [8]. The license must be explicitly permit distribution of software built from modified source code. The license may require derived works to carry a different name or version number from the original software.

## 4.5. No discrimination against individual / group

The license shall not be discriminate against any individual or group [8].

## 4.6. No discrimination against fields of endeavour

The license must not restrict anyone from making use of the program in a specific field of endeavour [8]. For example, it may not restrict the program from being used in a business, or from being used for generic research.

## 4.7. Distribution of license

The rights attached to the program must apply to all to whom the program is redistributed without the need for execution of an additional license by those parties [8].

## 4.8. License must not be specific only to a product

The rights attached to the program must not depend on the program's being part of a particular software distribution [8]. If the program is extracted from that distribution and used or distributed within the terms of the program's license, all parties to whom the program is redistributed should have the same rights as those that are granted in conjunction with the original software distribution.

## 4.9. License must not be restricting other software

The license must not be place restrictions on other software that is distributed along with the licensed software / application [8].

## 4.10. License should be Technological and Neutral

No provision of the license may be predicated on any individual or group of people's in technology or style of interface [8].

## 5. CATEGORICAL SUPPORTIVE TOOLS

Supportive tools in open source environment can be classified broadly in many categories (e.g. autonomous creative tools, domain-specific tools, domain-independent process support tools, computer supported collaborative tools etc).

### 5.1. Research tools

The research tools help the user in process of researching & improving the knowledge by disseminating the work previously done in the field [6, 17]. The search engines and the citation tracking tools fall into this category.

### 5.2. Computer supported collaborative tools

The CSCW tools allow the user to produce creative work using shared collaboration between multiple users [6, 15]. The collaboration becomes a need for a project involving multiple participants. The tool such as shared documents, shared editors and shared spreadsheets are the some examples of this category.

### 5.3. Domain-specific tools

These kinds of tools which allows the user to work at higher abstraction levels. Such kinds of tools are GIMP (GNU Image Manipulation Program) fall into this category and GIMP is an open source platform for image manipulations [6]. This tool allows the user to work with the images at higher abstraction levels such as photo retouching without paying attention to the fine details such as pixel setting etc.

### 5.4. Domain-Independent tools

Domain independent process support tools are the tools which support the process of creative works. The tool such as versioning systems like (SVN and CVS) that helps the user to check the alternative possibilities for the creative works without losing data [6].

### 5.5. Autonomous creative tools

The tools which facilitate the user to produce create works little or no input from the user. These kinds of tools are explicitly created to transform the thoughts of the user into creative forms of work [6, 18]. The examples of autonomous tools are AARON software programming.

## 6. PROCESS LIFE CYCLE

In development cycle, the implementation phase is consisting on six sub-phases [9];

1) Writing code and submitting to the OSS community for review.

2) Strength of OSS is the independent, prompt peer review.

3) Negative implications of breaking the build ensure that contributions are tested carefully before being committed.

4) Code contributions may be included in the development release within a short time of having been submitted. This rapid implementation being a significant motivator for developers.

5) Large number of potential debuggers on different platforms and system configuration ensures bugs are found and fixed quickly.

6)  Relatively stable debugged production version of the system is released.

Open source system development life cycle is totally in contrast with strategic planning. The haphazard principle of individual developers perceiving and itch worth scratching is totally suspended by corporate firms considering how to best gain competitive advantage from open source. For example, Red-Hat has published an architectural roadmap that details its plans to move open source systems towards middleware tools [10].

**Table 1.** Characterization in OS Systems

| Process | OS-System Phases | Product | OS-System Phases |
|---|---|---|---|
| Development life cycle | 1) **Planning-** An itch worth scratching. <br> 2) **Analysis-** Part of the conventional agreement upon knowledge in software development. <br> 3) **Design-** Firmly based on principles of modularity to accomplish separation of concerns. <br> 4) **Implementation-** A loop of (coding, review, pre-commit tests, development release, parallel debugging, production release ) <br><br> *** Often the planning, analysis and designing phase are done by an individual/core group, who serves as"*tail-light to follow*" in bazaar. | Domains | Horizontal Infrastructures (e.g. Operating systems, Utilities, Compliers, DBMS, Web and Print servers etc) |
| | | Primary Business Strategies | 1)  Value –Added-Service-Enabling. <br> 2)  Loss-Leader/Market-Creating. |
| | | Supporting | Fairly haphazard's-much reliable on email list/ bulletin boards, or on support provided by specialized software firms. |
| | | licensing | 1)  GPL, LGPL, Artistic License, BSD and emergence of commercially oriented MPL. <br> 2)  *"Viral"* term used in relation to licensing. |

Other proprietary companies have also seen the strategic potential of open source to alter the competitive forces at play in their industry, perhaps to grow market share or undermine competition. For example, IBM is a strong supporter of Linux systems because it erodes the profitability of the operating systems market and adversely affects competitors like Sun & Microsoft [10].

## 7. TRIALS FOR COMMUNITIES

Basically software architecture and functionality are governed by a community consisting of developers who can commit code to the authorized version of the software. When joining such a community contributors adhere to a social *"joining script"*, they put in considerably more technical and sophisticated work than the average contributor [11, 17].  After being admitted into the community, the new developers specialize in mundane software development tasks before moving on to more complex and technically difficult work. We think that there are eight main challenges that are faced by the open source community in making the open source systems [12]. In order to make efficient open source systems more useful in the computing world, theses problems should be addressed properly;

- **Disparity in work & benefits:** Using the application based on open source often requires additional work from individuals who do not perceive direct from the use of application.

- **Critical dilemma problems:** The developers' may not enlist the critical mass of users required to be useful, or can fail because it is never to any one individual`s advantage to use it.

- **Disruption of social processes:** The development community can lead to activity that violates social values, threatens existing political structures or may be de-motivates the users crucial to its success.

- **Exception handling:** Sometimes the developers may not accommodate the wide range of exception handling and improvisation that characterizes much group activity.

- **Unobtrusive accessibility:** Features that support group processes are used relatively infrequently, requiring unobtrusive accessibility and integration with more heavily used features.

- **Difficulty of evaluation:** Almost insurmountable obstacles to meaningful, a generalized analysis and evaluation of development community prevent us from learning from past experiences.

- **Failure of intuitions:** Intuition in product development environment is especially poor for multiuser applications, resulting in inefficient management decision and error prone designs.

- **Adoption process:** The development process requires more careful implementation in the workplace than product developers have confronted.

## 8. EXPANSIONS BY DISTINCT ROLES & INCUBATION SOLUTIONS

In considering the open source system projects, make it a point to understand the community as you familiarizes yourself with the code. One of the proposed improvements in the open source world for a better collaboration is to define the working process clear and there should be some dedicated roles in the environment like describe in the figure1;

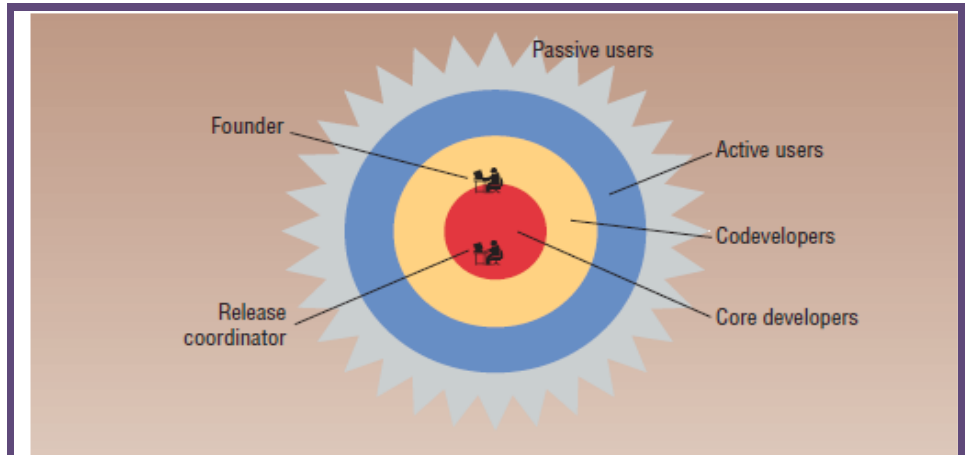

**Figure 1.** OS community *(like an onion with distinct roles for developers and leaders)* [12]

There should clear distinct roles for developers, leaders, and users. As shown in the figure 1 that in every domain there are dedicated roles that are responsible to perform their tasks. This proposed barometer in the development of open source systems shapes the size & tasks. As we know that the Apache and Eclipse both are the open source technologies by given the example here like a case study we propose an incubation solution for the effective process of development in open source environment [13, 18]. A two-stage incubation model based on analysis of Apache & Eclipse processes might help other organizations build their own incubation process and better manage risks. Two leading OSS communities, Eclipse and Apache, are successfully applying incubation processes.

**Apache Incubation:**

The incubator is the entry path into the ASF (Apache Software Foundation) for complicated projects and code bases wishing to become part of the Foundation's efforts and quality of work. The main responsibilities of incubator were;

a) To filter proposals and helps to create creative projects

b) Evaluate a project's maturity promote the creation of a community that shares the ASF principles, including meritocracy as one of central element.

Two types of projects can accommodate: first subprojects that have finalized the incubation period and are added into an already existing top-level project, and second the top-level projects that shape the main lines of the foundation's evolution and can sponsor other incubated subprojects [13]. The Incubator PMC (Project Management Committee) supervises the incubation processes. The PMC accepts projects and provides the technical and administrative support required; it regularly reviews incubated projects, proposing them for termination, continuation, or escalation [13]. In addition to the PMC, some roles are defined per project. Candidates put forward project proposals. The champion is a foundation officer or member who helps candidates make their initial submission to a sponsor [13].

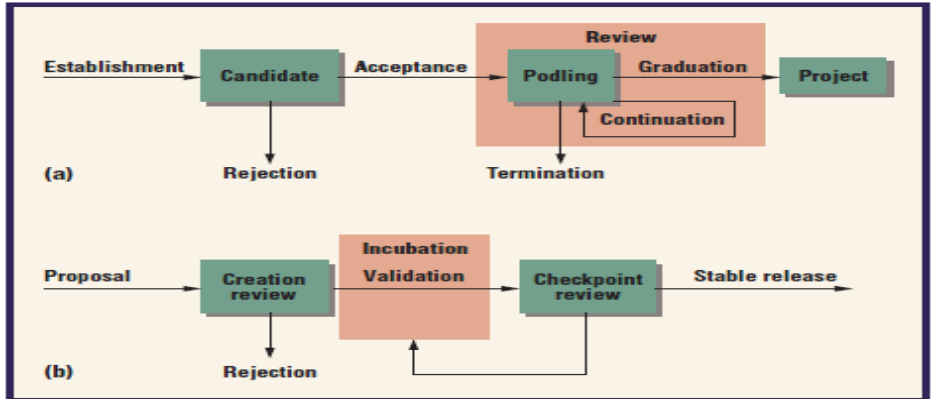

**Figure 2.** (a) Apache phases of establishment, acceptance, and review.
(b) Eclipse incubation / validation process [13].

**Eclipse Incubation:**

Here it is EMO (Eclipse Model Operations) that works like the PMC in the apache case. In eclipse, projects are incubated within the top-level technology project, concentrating in this area all the project creation experience. An eclipse high-level structure differentiates between standard and top-level projects. Each top-level project has several projects [13, 20]. Each standard project organization consists of the project lead and the development team, who are respectively responsible for planning the project and defining the technical architecture [13]. In the eclipse incubator evaluates the proposal into check that;

a) The community is active and the project is operating fully in the open using open source rules of engagement.

b) The project adopts eclipse's philosophy and principles.

c) An in-depth review of the project's technical architecture takes into accounts its dependencies and interactions with other projects.

## 9. CONCLUSION

We need better understanding of collaboration in the open source development process. Collaboration in the open source project development has been improving these days but however there is more scope for these improvements. Adapting the incubation process can be used by companies to promote bottom-up innovation inside the corporation. Incubation might provide a framework in which individuals can contribute their ideas and gather help and feedback in an informal way, while at the same time providing a means for selecting which ideas have enough potential to be further developed. The open source communities are in constant evolution: at the time of this writing, the eclipse community is reviewing its incubation process to make it more agile, Apache is introducing new initiatives trying to foster innovation, and incubation is mandatory in both communities [13, 19]. In order to improvement the collaboration in the open source environment both the development community and the people should be more creative.

Of course, in today's technological era, the perspective of information systems based on the open source technology has to be changed towards the non profitable settings such as education and community informatics etc. Open source increase the horizons of creative thought process as users can use the free tools and can change the binaries of the software to suits to their needs.

This facilitates growth of open source community which can engage audience from different fields and share the knowledge for producing new versions and software. This will also help in making the large projects with the limited resources using a good collaboration between the people from the different technicalities.

## ACKNOWLEDGEMENTS

I thank the anonymous reviewers for invaluable guidance on improving this article. Iam grateful to Professor.Eric Monteiro of Norwegian University of Science and Technology (NTNU) for helping to locate information & material for this article. Finally, I am thankful to my parents for their all-time love, financial support and their prayers for me.

## REFERENCES

[1]     Gvon Krogh and E von Hippel (2003)" *Special issue on open   source software development"*, *Research     Policy*, 32(7):1149-1157, 2003.

[2]     Dam, K.W., (1995) "Some economic considerations in the intellectual property protection of software". *Journal of Legal Studies* 24 (2), 321–377.

[3]     Perens.B.,(1998)"The       Open       Source       Definition".       Available       online       at: http://www.perens.com/articles/osd.html [Retrieved on 4th December, 2010].

[4]     Raymond, E., (1999) *The Cathedral and the Bazaar: Musings on Linux & Open Source by an Accidental Revolutionary*. O'Reilly Publishers, Sebastopol, CA. USA.

[5]     The       SEI       Report–USA       (2007).       Available       online       at http://www.web.archieve.org/web/20070903115947.htm [Retrieved on 9th December 2010].

[6]     Cusumano, M. A., 1991, *Japan's Software Factories: A Challenge to U.S. Management* (Oxford University Press, New York, NY).

[7]     Sahay and Robey *"Transforming work through information technology: a comparative   case study of geographic information systems in county government, Information Systems Research"*, 7(1):63-92, 1996.

[8]     The Open Source Initiative (OSI), San Francisco CA 9411, United States of America. Available online at:  http://www.opensource.org/docs/osd [Retrieved on 11th December 2010].

[9]     Feller, J. and Fitzgerald, B. (2002) *Understanding Open Source Software Development*. Addison-Wesley; UK.

[10]    Brian.F (2006)   *"The Transformation of Open Source Software"* , Irish Software Engineering Research Centre, University of Limerick, Ireland. MIS Quarterly, Vol. 30, No. 3, 2006.

[11]    Jonathan Grudin (1989) "Why groupware applications fail: Problems in design and evaluation. Office: Technology and People", 4(3):245-264, June 1989.

[12]    Kevin.C; James.H (2006) "Assessing the Health of Open Source Communities". IT-Systems Perspectives. Syracuse University USA.

[13]    Juan C. Hugo A. Parada G.,(2007) *"Apache and Eclipse: Comparing Open Source Project Incubators"*, IEEE Software   0740-7459/07. 2007- USA.

[14]    D. Weiss and C.T.R. Lai (1999) *Software Product Line Engineering: A Family Based Software Development*. Addison-Wesley, USA.

[15]    Gibbons, R. and Waldman, M., 1999, 'Careers in Organizations: Theory and Evidence', in O. Ashenfelter and D. Card (eds), *Handbook of Labour Economics*, vol. 3B, Chapter 36 (North Holland, New York).USA

[16]  Wayner, P., 2000, *Free for All: How Linux and the Free Software Movement Undercut the High-Tech Titans* (Harper Collins, New York).USA

[17]  Von Hippel, E., 1988, *The Sources of Invention* (Oxford University Press, New York).

[18]  Rosenberg, N., 1976, *Perspectives on Technology* (Cambridge University Press, Cambridge).UK

[19]  Merton, R. K., 1973, *The Sociology of Science: Theoretical and Empirical Investigation* (University of Chicago Press, Chicago).USA.

[20]  Holmstr6m, B., 1999, 'Managerial Incentive Problems: A Dynamic Perspective', *Review of Economic Studies*, 66, pp. 169-182.

**Author:**

**Murtaza Hussain Shaikh**, has earned Master`s in (Information Systems Engineering) from Norwegian University of Science & Technology (NTNU), Norway and another Master`s in (Science and Technology Policy) from Mehran University of Engineering & Technology (MUET), Pakistan. He got his Bachelor`s degree in (Software Engineering) from University of Sindh, Pakistan and another Bachelor in (Arts &Economics) from Allama Iqbal Open University, Pakistan. His current research interests include IS Security, Technological Law and Policy Making of Computer Organizations, Systems operations, and Cyber Security issues. He is a Research Scientist in leading research organization of Pakistan and teaching experiences in different Pakistan`s Universities and Colleges. He has authored and co-authored research articles in IEEE and International Journals.

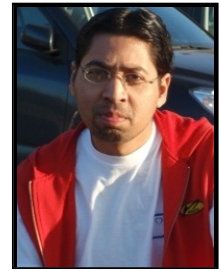